\documentclass[preprint,authoryear,10pt,letterpaper]{elsarticle}
\usepackage[top=0.75in, bottom=0.75in, left=0.75in, right=0.75in]{geometry}

\usepackage{palatino}
\usepackage{mathpazo}
\usepackage{amsmath,amsthm}
\usepackage{amssymb}
\usepackage{amsfonts}
\usepackage{graphicx}
\usepackage{subfigure}
\graphicspath{{figures/}}
\usepackage{algorithm}
\usepackage{algorithmic}
\usepackage{tabularx,booktabs}

\usepackage{tablefootnote}

\usepackage{xcolor}
\definecolor{darkblue}{rgb}{0.0,0.5,0.5}
\definecolor{blue}{rgb}{0.0,0.5,0.68}
\usepackage[colorlinks]{hyperref}
\hypersetup{colorlinks,breaklinks,linkcolor=darkblue,urlcolor=darkblue,anchorcolor=darkblue,citecolor=darkblue}
\usepackage{booktabs,caption}
\usepackage{multirow}
\usepackage{lscape}
\usepackage{epstopdf}
\usepackage{lineno}
\usepackage{subfigure}
\usepackage{multirow}
\usepackage{makecell}

\usepackage{microtype}
\usepackage{lineno}
\journal{ }
\makeatletter
\def\ps@pprintTitle{%
   \let\@oddhead\@empty
   \let\@evenhead\@empty
   \let\@oddfoot\@empty
   \let\@evenfoot\@oddfoot
}
\makeatother

\makeatother

\begin{document}

\begin{frontmatter}



\title{Routine Pattern Discovery and Anomaly Detection in Individual Travel Behavior}

\author[label1,label2]{Lijun Sun\corref{cor1}}
\ead{lijun.sun@mcgill.ca}

\author[label1]{Xinyu Chen}

\author[label3]{Zhaocheng He}

\author[label1,label2]{Luis F. Miranda-Moreno}

\address[label1]{Department of Civil Engineering, McGill University, Montreal, QC H3A 0C3, Canada}
\address[label2]{Interuniversity Research Centre on Enterprise Networks, Logistics and Transportation (CIRRELT)}
\address[label3]{School of Intelligent Systems Engineering, Sun Yat-Sen University, Guangzhou, Guangdong 510006, China}

\cortext[cor1]{Corresponding author. Address: 817 Sherbrooke Street West, Macdonald Engineering Building, Montreal, Quebec H3A 0C3, Canada}

\begin{abstract}
Discovering patterns and detecting anomalies in individual travel behavior is a crucial problem in both research and practice. In this paper, we address this problem by building a probabilistic framework to model individual spatiotemporal travel behavior data (e.g., trip records and trajectory data). We develop a two-dimensional latent Dirichlet allocation (LDA) model to characterize the generative mechanism of spatiotemporal trip records of each traveler. This model introduces two separate factor matrices for the spatial dimension and the temporal dimension, respectively, and use a two-dimensional core structure at the individual level to effectively model the joint interactions and complex dependencies. This model can efficiently summarize travel behavior patterns on both spatial and temporal dimensions from very sparse trip sequences in an unsupervised way. In this way, complex travel behavior can be modeled as a mixture of representative and interpretable spatiotemporal patterns. By applying the trained model on future/unseen spatiotemporal records of a traveler, we can detect her behavior anomalies by scoring those observations using perplexity. We demonstrate the effectiveness of the proposed modeling framework on a real-world license plate recognition (LPR) data set. The results confirm the advantage of statistical learning methods in modeling sparse individual travel behavior data. This type of pattern discovery and anomaly detection applications can provide useful insights for traffic monitoring, law enforcement, and individual travel behavior profiling.
\end{abstract}


\end{frontmatter}

\section{Introduction}

With the rapid development of information and communications technology (ICT), large amounts of spatiotemporal data---such as GPS trajectory data, license plate recognition (LPR) data, call detailed records (CDR), and transit smart card transactions---are generated continuously through individuals' mobility and travel activities. The high spatial/temporal resolution of these data sets sheds new lights on advancing our understanding of human mobility and travel behavior, which have been shown to be highly consistent and predictable over time in previous studies.

The intrinsic regularity of travel behavior allows planners and practitioners to design better transport systems and services based on simplified and aggregated patterns \citep{kitamura2006routine,schonfelder2010urban,sun2013understanding}. However, given the increasing amount of individual-based spatiotemporal data, we may observe anomalous behavior frequently in contrary to individual regularity. For instance, a common assumption of smart card data is to consider each ID a unique user, while in reality, two users may share one card for their daily transit use. The same applies to vehicle usage, for which we consider license plate number a proxy to a unique driver, while a car in a household can be shared among many members. When this happens, we may observe anomalous behaviors with regard to the regularity and stability of individual travel behavior. Moreover, in the long term, an individual's travel pattern may also change over time, such as changing jobs and moving to a new address. For example, \citet{zhao2018detecting} studied pattern changes in individual travel behavior using long-term smart card transactions. The authors define travel pattern changes as ``abrupt, substantial, and persistent changes in the underlying travel patterns'' and apply a moving kernel to measure the degree of changes over time for frequency, spatial, and temporal features, respectively.

Detecting such anomalies in travel behavior is essential to many transportation planning, traffic operation, and law enforcement applications. At a collective level, the anomaly may come from intrinsic variations resulted from incidents, events, policy implementation, and infrastructure constraints. For instance, travel behavior of students may change substantially from a spring semester to the following summer vacation. Transport agencies should take this collective behavior shift into consideration in their daily traffic management practices. At an individual level, the anomaly may arise from factors such as change of home/work location, road space rationing policy, or even sharing one's car with a friend. However, given the high complexity and large volume in emerging spatiotemporal behavioral data, it is infeasible to examine the data manually to identify the behavior anomalies mentioned above.

This paper attempts to provide a unified framework for characterizing meaningful patterns in travel behavior and then detecting anomalies based on the learned patterns. We define individual behavior anomalies as cases where it is difficult to predict one's current/future activity given her past travel behavior. The contribution of this paper is two-fold. First, we develop a generative model for individual spatiotemporal travel records. Specifically, we extend the latent Dirichlet allocation (LDA) model \citep{blei2003latent} to generate spatiotemporal records and use a two-dimensional probability distribution to model a traveler's topic distribution. The model can (1) uncover meaningful spatiotemporal travel behavior patterns by sharing information from a large number of users and (2) summarize an individual's complex mobility data into a low-dimensional latent space, which not only provides a powerful way to predict one's future mobility patterns but also allows us to quantify the similarity between users and cluster users. This knowledge is critical to many smart transportation applications such as trip planning and car sharing. The compact representation obtained for each individual can also be used in other applications such as clustering travelers with similar behavior patterns. Second, on top of the generative mobility model, we propose a probabilistic framework for detecting anomalies in individual travel behavior using perplexity as a scoring function. A similar approach has been developed in \citet{xiong2011group} and \citet{yu2015glad}, but in this paper we focus on individual behavior: we measure the degree of behavior anomaly for each individual traveler by using the ``predictability'' of future mobility records under the trained model. A high perplexity indicates that an individual's future behavior cannot be well reconstructed by her past behavior. And a low perplexity score suggests that the model can well predict an individual's future travel behavior. To verify the effectiveness of this framework, we conduct an empirical analysis based on a large-scale LPR data set collected from Guangzhou, China. Note that the general framework can be applied to a variety of mobility data with spatial and temporal information encoded (e.g., smart card data and GPS data).

The remainder of this paper is structured as follows. In Section~\ref{sec:literature}, we review relevant literature on spatiotemporal mobility modeling at both individual and collective levels, and recent work on anomaly detection in travel behavior. In Section~\ref{sec:data}, we introduce the LPR data used in this study. Section~\ref{sec:model} presents the key LDA-based model for spatiotemporal mobility modeling, including the inference algorithm and the anomaly detection framework. In Section~\ref{sec:experiments}, we conduct extensive numerical experiments to demonstrate the effectiveness of this framework. Section~\ref{sec:conclusion} concludes this study and provides some future research directions.

\section{Literature Review} \label{sec:literature}

There exists a large body of literature on modeling routine patterns and regularity in human mobility and travel behavior. Most traditional analyses are based on travel survey data together with the discrete choice modeling framework  \citep[e.g.,][]{hanson1988systematic,bhat2000comprehensive,bowman2001activity,axhausen2002observing,buliung2008exploring}. With the recent development of ICT, collecting large-scale and high-resolution spatiotemporal data sets at the individual level has become much easier and less costly. Examples of such data sets include call detailed records (CDR) \citep[e.g.,][]{gonzalez2008understanding,ahas2010daily,widhalm2015discovering}, location-based social networks (LBSN) \citep[e.g.,][]{hasan2014urban}, transit smart card transactions \citep[e.g.,][]{hasan2013spatiotemporal,sun2016understanding,zhao2018detecting}, and GPS trajectories \citep[e.g.,][]{zheng2012unsupervised}, to name but a few.

The availability of these data sets has provided us with unprecedented opportunities to study routine patterns in individual travel behavior. This topic has attracted researchers from various fields beyond transportation, such as statistical physics and computer science. In the field of statistical physics, some landmark works have been conducted to understand and measure the regularity rooted in human mobility with an interdisciplinary approach. For example, \citet{gonzalez2008understanding} conducted the first statistical physics study on the regularity of human mobility based on a large-scale CDR data set and introduced the radius of gyration measure to quantify human mobility. Later, \citet{song2010limits} studied the limits of predictability in human mobility from an information theory perspective using the same CDR data set. \citet{schneider2013unravelling} identified the motifs rooted in human mobility patterns using large-scale CDR data. The authors found that each individual in general has a specific characteristic motif which is stable over months. \citet{sun2013understanding} studied individual spatiotemporal regularity using smart card data and revealed the social phenomenon of ``familiar strangers'', which further confirmed the strong regularity and daily circadian rhythms in individual mobility. We refer interested readers to \citet{barbosa2018human} for a comprehensive review of recent works in statistical physics on human mobility models and applications.

Although statistical physics models have brought substantial advances in travel behavior research, they are still limited in dealing with the high dimensionality, complexity, sparsity and missing value problems in large-scale spatiotemporal data sets. These challenges have attracted more and more researchers from computer science. As a result, machine learning models have been increasingly used to model individual mobility patterns. Essentially, these works mainly focus on developing and applying data-driven methods---such as principal component analysis (PCA), matrix/tensor factorization, hidden Markov model (HMM), and probabilistic topic model---to identify latent spatiotemporal patterns within the data. The pattern recognition process not only allows one to cluster individuals based on their activity pattern but also provides a generative mechanism to predict future activities. However, the machine learning approaches are often overlooked by the travel behavior research community. In the following, we review some signature works on (1) modeling spatiotemporal travel behavior using machine learning models, and (2) applying statistical learning models to detect anomalies in travel behavior.

\subsection{Machine Learning for Spatiotemporal Travel Behavior Modeling}

As mentioned, there has been increasing interest in applying statistical learning models to capture the novelty and patterns in human mobility data sets. \citet{eagle2009eigenbehaviors} are among the first data-driven studies on human behavior patterns using smart phone data. In the Reality Mining project, the authors collected behavior data using mobile phones from 97 users over 16 months, and then applied PCA on temporal activity data to extract primary routines in daily behavior. Although the data set is not large, it still characterizes meaningful behavior patterns, which are referred to as ``eigenbehaviors'' in the paper. Based on the same data set, \citet{farrahi2011discovering} developed a topic model using the combination of activity labels (e.g., H-home, W-work, O-other) and timestamps as input. This topic modeling approach can summarize spatiotemporal data into certain routine patterns, and further identify users with similar behavior patterns. Similarly, \citet{jiang2012clustering} applied PCA on an activity survey data set collected from Chicago. Based on the projection vectors, the authors applied the k-means clustering algorithm to group people into different categories given their activity patterns. \citet{hasan2014urban} applied topic model on geo-location data extracted from Twitter to detect patterns in individual activity choices. When combined with traditional travel survey data, the model could serve as an activity generation simulator. A limitation of this model is that the authors define a word as a joint combination of three attributes (day of week, time of day and activity type), and thus it creates a large set of vocabulary, and the complex correlations among different attributes are essentially ignored. In a recently work, \citet{goulet2016inferring} also applied PCA on a large-scale multi-week transit activities (smart card) data set and revealed 11 clusters of transit users based on their travel sequences.

To better model the interaction between the spatial and the temporal dimensions, recent research has started looking at the joint distribution spatiotemporal observations. For example, \citet{mcinerney2013modelling} proposed a Bayesian non-parametric topic model to characterize the generative mechanism of the joint observation of day of week, time of day, and location. The proposed generative process has two main features: (1) temporal factors for day and time of day are shared across the population, and (2) each individual has his/her own spatial distribution. A Dirichlet process (DP) is imposed on topic distribution and this provides a non-parametric way to adjust model complexity given the input data. \citet{fan2016collaborative} essentially applied the same topic model (except removing the DP prior on topic distribution) on a large-scale mobile phone data. In order to infer missing location and complete individual trajectories, the authors further integrated a HMM layer to predict user's location when it is not observed in the raw data. \citet{huai2014toward} also employed a Bayesian HMM framework to learn meaningful and dynamic patterns from spatiotemporal trajectories. \citet{zheng2012unsupervised} developed a variant of topic model which generates both spatial and temporal observations jointly. A Gaussian mixture layer is introduced to model the temporal distribution of activities. In a later extension, \citet{zheng2013effective} developed a collaborative filtering technique by factorizing the sparse spatiotemporal matrix into meaningful latent patterns. \cite{baratchi2014hierarchical} propose a hidden semi-Markov-based model to understand the spatiotemporal behavior of mobile entities. In this work, a hierarchical state structure is used to capture both activity location and the path connecting two consecutive activities. \citet{widhalm2015discovering} combined two mobility data sets and proposed to use an undirected relational Markov network to infer urban activities with CDR data. In a recent work, \citet{qin2018spatio} developed a new Spatio-Temporal Routine Mining Model (STRMM), which is essentially a topic model on daily behaviors with factors defined on each timespan.

Besides uncovering spatiotemporal mobility patterns, most of these probabilistic models also characterizes the generative process for individual travel behavior, which can be further used for prediction tasks such as generating future travel behavior/mobility patterns. We refer to this research question as synthetic activities and synthetic travel demand. Along this track, some studies have considered predicting human mobility patterns as a supervised learning problem, e.g., the prediction of next location and the prediction of next activity/trip. For example, \citet{zhao2018individual} built a Bayesian n-gram model to predict the next trip (time, origin, and destination) given the previous one. The model was tested on a smart card data set collected from London, UK and showed good performance. Notably, \citet{yin2018generative} proposed a full framework for individual activity synthesis based on mobile phone data. The authors used the input-output (IO) HMM architecture to jointly generate all variables in travel activities---including purpose, mode, destination, and duration of stay, etc.

At the collective level, there are also some works on modeling aggregated travel demand as a whole. For example, \citet{fan2014cityspectrum} applied non-negative tensor factorization on a large GPS log data set to uncover latent factors for different regions, at different times, and on different days. The decomposed tensor could be used for site recommendation and mobility flow simulation purposes. \citet{sun2016understanding} proposed to use a probabilistic tensor factorization model to reconstruct a fourth-order (time of day$\times$passenger type$\times$origin zone$\times$destination zone) tensor for collective mobility. The algorithm is able to characterize the interdependencies and correlations on both spatial (origin/destination) and temporal dimensions.

\subsection{Anomaly Detection in Mobility and Travel Behavior}

Despite recognizing patterns in the data, statistical learning models also provide a  probabilistic scheme to detect how ``abnormal'' an activity might be. Most methods for anomaly detection follow a similarity and clustering-based approach \citep{chandola2009anomaly}, with a specific purpose of identifying abnormal activities from a population. In detecting anomalies in spatiotemporal data sets, we often consider three sub problems: 1) spatiotemporal outlier detection, 2) spatiotemporal outlier tracking, and 3) trajectory outlier detection \citep{gupta2014outlier}. In terms of mobility and travel behavior, we are more interested in detecting anomalous activities of an individual with regard to his/her past travel activities. Although this is an exciting and important research question, the literature on detecting anomalies in travel behavior is much limited compared with the large body on pattern discovery. However, in fact, all probabilistic generative models can be adapted in a certain way to fulfill the anomaly detection tasks as long as we have a well-defined scoring system. For example, \citet{xiong2011group} proposed to use perplexity---a measure which is often used for model selection---to evaluate the degree of anomalous in a data set.

In the context of urban mobility, \cite{zhang2018detecting} defined anomaly as the abnormal movement of crowds and incident/accident and proposed a two-step method for urban anomaly detection. The authors also gave a comprehensive review about detecting urban anomalies. We refer interested readers to this paper and the references therein for more works about urban anomaly detection. In term of individual travel behavior, most anomaly detection techniques are similarity and cluster/classification based, with the goal of identifying abnormal individuals/travelers from a population. For example, \citet{zhao2017spatio} data mining techniques on massive smart card data and detected abnormal users. In this paper, the examples of anomalous users are those who abuse the metro system, such as homeless people and thefts who travel back and forth for long time.  \citet{witayangkurn2013anomalous} developed an HMM-based framework to detect abnormal urban mobility based on number of unique users. \citet{du2016catch} developed a feature-engineering approach to identify important features separating Pickpocket Suspects from normal travellers from transit smart card data.

To the best of our knowledge, only a few previous studies have addressed the anomaly detection problem at the individual level. For example, \citet{shih2016personal} studied individual mobility patterns and use patterns as input for anomaly detection. The authors motivate the research question by considering patients suffering from Alzheimer's disease and developed an trajectory-based algorithm to detect abnormal spatiotemporal patterns. In this paper, we focus on detecting anomalies in one's travel behavior against his/her past mobility records. In doing so, we proposed an integrated probabilistic framework which can perform pattern recognition and anomaly detection of individual travel behavior simulaneously.

\section{License Plate Recognition Data} \label{sec:data}

With the advances in video camera and image processing techniques, the LPR system has become an efficient and effective way to capture real-time vehicle location information \citep{herrera2010evaluation}. Since its development, the LPR system has played an important role in modern intelligent transportation systems (ITS) and been widely applied in various practical urban/highway transportation applications, such as traffic monitoring/management, road pricing, and law enforcement. In the literature, LPR data has been used in addressing many research/practice questions, including link travel time estimation \citep{bertini2005validating,kazagli2013estimation}, queue length estimation \citep{zhan2015lane}, speed profile estimation \citep{mo2017speed}, and travel pattern clustering \citep{chen2017clustering}, OD matrix and path flow reconstruction \citep{castillo2008trip}, missing data imputation \citep{zhang2019missing}, to mention but a few.

\begin{table}[htbp]
  \centering
  \caption{Example records of LPR data in Guangzhou}
    \begin{tabular}{lrrr}
    \toprule
    Vehicle ID & Location ID &   Direction & Time stamp \\
    \midrule
    1c8a5aef8e8785243173283c04a558fb    & 17060 & 3     & 03/01/2017 07:19:11 \\
    1c8a5aef8e8785243173283c04a558fb     & 17005 & 2     & 03/01/2017 07:23:56 \\
    733e2241fc6d380375a1dba820fecc72     & 17054 & 0     & 03/14/2017 20:35:24 \\
    733e2241fc6d380375a1dba820fecc72    & 17058 & 0     & 03/14/2017 20:48:35 \\
    9d213b84d6cda80e9be89baa1c5f54c7     & 17069 & 1     & 03/18/2017 19:52:08 \\
    9d213b84d6cda80e9be89baa1c5f54c7      & 17059 & 2     & 03/18/2017 20:14:56 \\
    9d213b84d6cda80e9be89baa1c5f54c7    & 17037 & 1     & 03/18/2017 21:08:14 \\
    1c8a5aef8e8785243173283c04a558fb      & 17060 & 3     & 03/01/2017 07:19:11 \\
    1c8a5aef8e8785243173283c04a558fb     & 17005 & 2     & 03/01/2017 07:23:56 \\
    \bottomrule
    \end{tabular}%
  \label{tab:lpr}%
\end{table}

\begin{figure}[!ht]
\centering
\includegraphics[width=5cm]{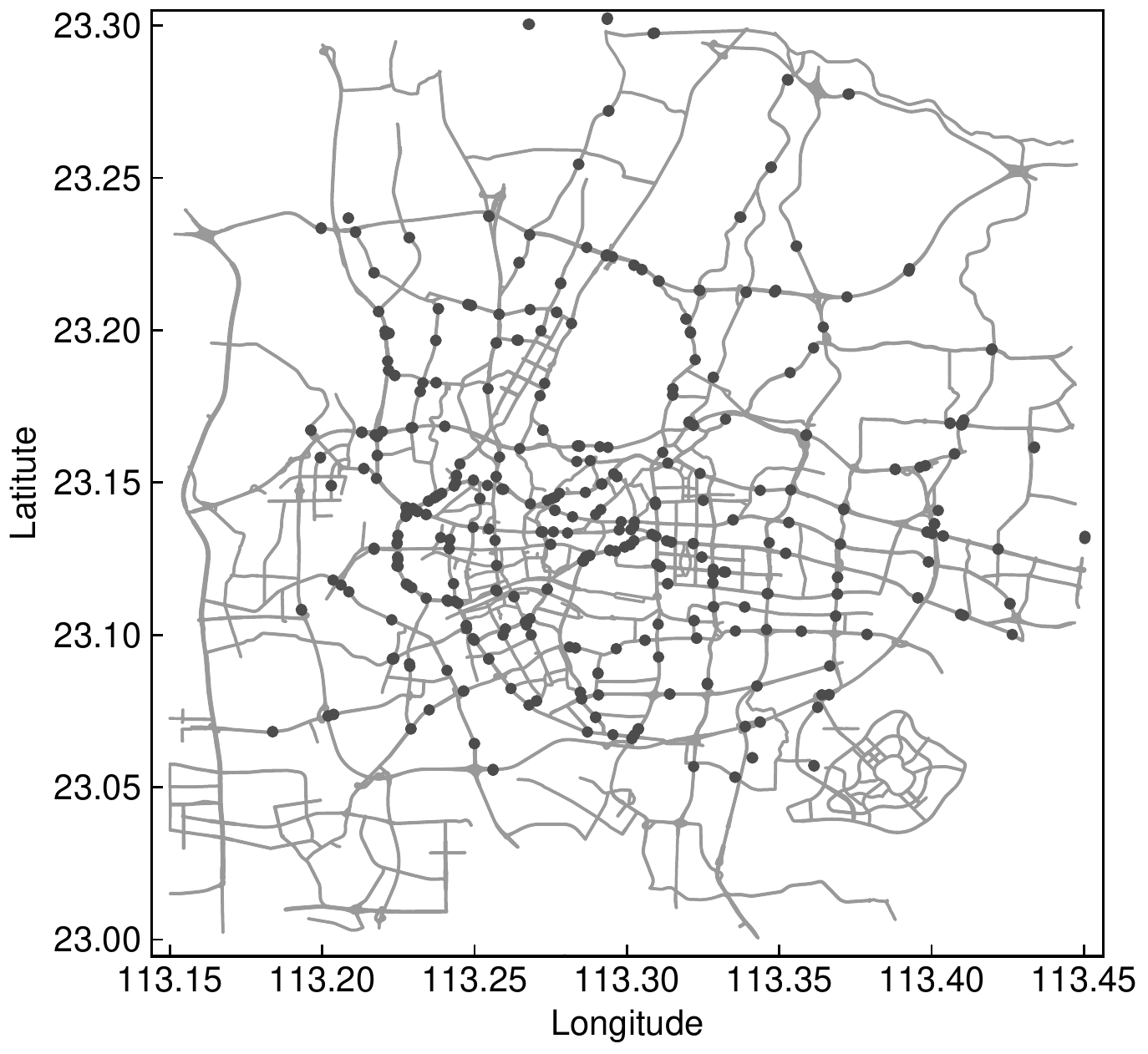}\hspace{.2cm} \includegraphics[width=5cm]{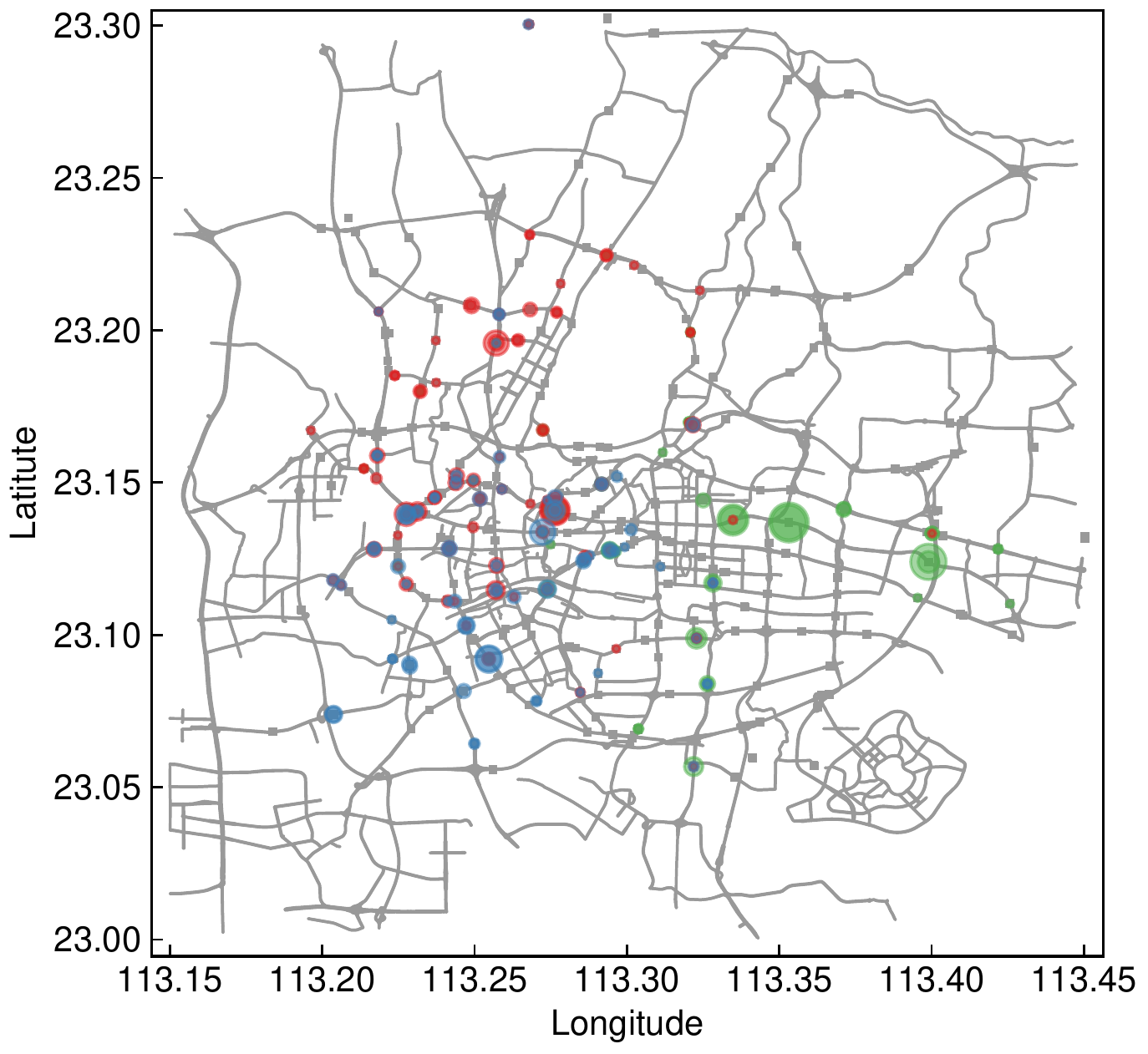}
\caption{License plate recognition data in Guangzhou: (left) detectors location in the metropolitan area of Guangzhou, and (right) detector visit frequency of three vehicles (in red, blue, and green).}
\label{fig:fig1}
\end{figure}

Our study is based on a four-week (March 1-28, 2017) LPR data set collected in Guangzhou, China. For each record, the LPR system registers vehicle ID (license plate number), location (gantry ID), vehicle driving direction, the color of the license plate, together with a timestamp. Table~\ref{tab:lpr} gives some example records of the LPR data set in Guangzhou (license plate number is anonymized). We combine location and vehicle driving direction as a new spatial feature, which is referred to as ``detector'' in the following of this paper. We consider each LPR record a tuple with three elements - $\left(u,t,s\right)$, where $u$, $t$ and $s$ represent vehicle ID, timestamp and detector ID (proxy of location), respectively. The left panel in Figure~\ref{fig:fig1} shows the location of all detectors in the metropolitan area of Guangzhou, and the right panel shows detector visiting frequency of three example travelers (vehicles IDs) in different colors. Due to privacy and confidentiality concerns, in this study we only have access to a small subset containing transactions from 2200 unique travelers/license plate numbers. The number of transactions in total is 2,928,452.

\section{A Generative Model for Spatiotemporal Data} \label{sec:model}

In this section, we present a generative model for the spatiotemporal LPR data, and then flagging behavior anomalies as records that are unlikely to be generated by the model. In detail, we use a hierarchical mixed membership model, like LDA, to characterize the data generative processes. As mentioned, the LPR data provides spatiotemporal travel information for each vehicle ID. For simplicity, we treat each vehicle ID as a unique traveler. We denote by $S$ the total number of detectors on the spatial dimension ($S=463$ in the Guangzhou LPR data). On the temporal dimension, we group the timestamps (time of day) at an hourly interval, and thus the size of the temporal dimension is $T=24$. We denote by $$\mathbf{w}_u=\left\{ (w_{ui}^t,w_{ui}^s): i=1,\ldots,N_u,\  w_{ui}^t\in \left\{1\ldots T\right\},\  w_{ui}^s\in \left\{1\ldots S\right\} \right\}$$ the collection of all spatiotemporal records of traveler $u$, where $N_u$ is the total number of LPR records. With this definition, we consider a traveler as a ``bag of spatiotemporal points'' and make the analogy between ``document''---``vocabulary''  and ``traveler''---``spatiotemporal point''.

\subsection{Model Specification}

Although the analogy gives the idea to apply text models on the travel behavior data sets, a critical problem is to first formally define ``word'' and model the spatial and temporal dimensions jointly. A straightforward approach is to combine spatial and temporal features into a single dimension with a vocabulary size of $S\times T$, such as in \citet{hasan2014urban}. However, there are two problems with this approach: on the one hand, a large set of vocabulary is introduced; on the other hand, the clear interdependency among the linked words (e.g., same time but different detector, or same detector but different time) is essentially ignored. As a result, each document will be modeled as a sparse vector without considering the explicit association between similar words (e.g., [Monday, 12pm, Eat] and [Tuesday, 12pm, Eat]: the only difference is the day of week).

A straight forward solution is to separately model the spatial and the temporal dimension. For example, \citet{matsubara2012fast} applied a multi-way LDA model to mine patterns for complex event data. Most existing models for multi-way LDA still follow the assumption of conditional independence (on different dimensions) given a selected topic. This approach indeed can explore the full distribution; however, in practice, even simple dependence structure across different dimensions/attributes may require a very large number of factors.

\begin{figure}[!ht]
\centering
\includegraphics[scale=0.3]{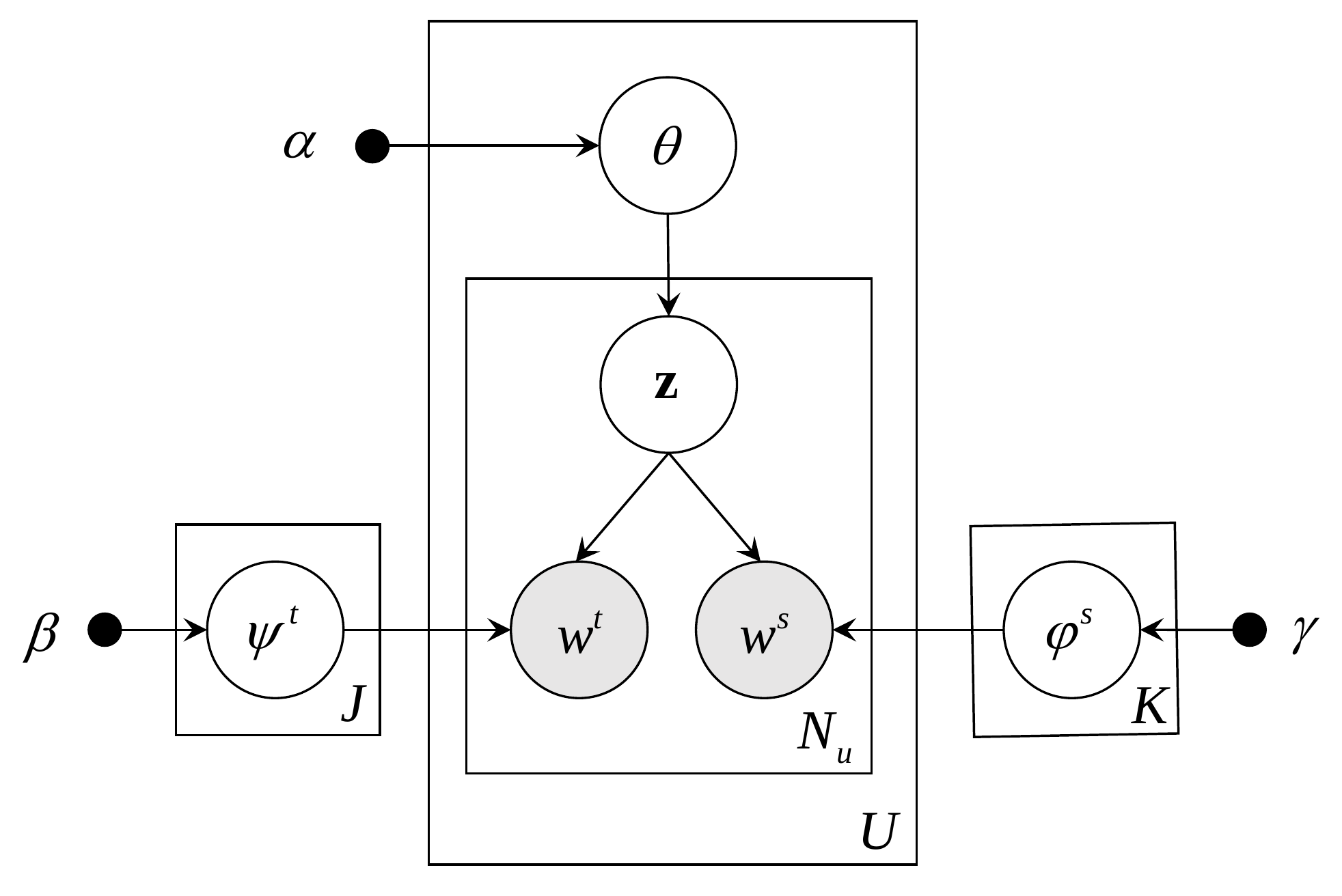}
\caption{Graphical model of the proposed spatiotemporal LDA model. Note here $\mathbf{z} = (z^t, z^s)$ represents a spatiotemporal topic combination. }
\label{fig:lda}
\end{figure}

To address this issue, in the following we introduce a two-dimensional spatiotemporal LDA model, which generates the spatial and temporal feature jointly while capturing their dependencies. We impose two properties in this generative model: (1) both spatial and temporal features are considered ``words'' and they are generated in a joint manner. Therefore, two sets of topic-word distributions are determined for spatial patterns and temporal patterns, respectively; (2) instead of a vector Dirichlet, an individual's topic distribution is defined on a two-dimensional simplex, with each cell capturing the degree of interaction between a spatial topic and a temporal topic. We show the graphical model in Figure~\ref{fig:lda}. Here, the individual topic distribution $\theta_u$ is defined on a two-dimensional probability simplex of $J\times K$, with each element $0\le\theta_{ujk}\le 1$ and $\sum\nolimits_{j=1}^J \sum\nolimits_{k=1}^K \theta_{ujk}= 1$. Therefore, $\theta_{ujk}$ actually characterize the degree of interaction between temporal topic $j$ and spatial topic $k$ for traveler $u$. These two properties give us a flexible representation to capture more variation/heterogeneity in the data with less number of topics (similar to Tucker decomposition, see \citet{sun2016understanding}). In particular, this model allows us to introduce different numbers of topics ($J$ and $K$ as hyperparameters) on both the temporal and spatial dimensions, respectively. This feature is advantageous in our case since the vocabulary size on the spatial and temporal dimensions ($S=463$ and $T=24$) vary substantially. Below we summarize the generative process (see Figure~\ref{fig:lda}):

\begin{itemize}
  \item draw topic distribution for each traveler $\theta_u \sim\text{Dirichlet}_{J\times K}(\alpha)$
  \item draw time distribution for each temporal topic $\psi_j\sim\text{Dirichlet}_{T}(\beta)$
  \item draw detector distribution for each spatial topic $\varphi_k\sim\text{Dirichlet}_{S}(\gamma)$
  \item for each traveler $u$, for each LPR record:
  \begin{itemize}
  \item draw a topic $\mathbf{z} = \left(z^t, z^s\right)\sim\text{Multinomial}_{J\times K}(\theta_u)$
  \item draw a word $w^{t}\sim\text{Multinomial}_{T}(\psi_{z^t})$
  \item draw a word $w^{s}\sim\text{Multinomial}_{S}(\varphi_{z^s})$
  \end{itemize}
\end{itemize}

In this model, $\alpha$, $\beta$, $\gamma$ are the Dirichlet priors, $\psi_{j}$ is the $j$-th temporal topic-word distribution, and $\varphi_{k}$ is the $k$-th spatial topic-word distribution. Note that the two-dimensional Dirichlet prior for individual topic distribution is the same as converting $\theta_u$ as a vector. The same applies to the two-dimensional multinomial distribution in sampling $\mathbf{z}=\left(z^t, z^s\right)$. In terms of methodology, the proposed model is a natural extension of LDA on spatiotemporal data.

\subsection{Model Inference}

In this section, we present a collapsed Gibbs sampling algorithm to infer the two-dimensional LDA model. First, we write the probability of observing a spatiotemporal record as:

\begin{equation}
P\left( {{w_{ui}^t},{w_{ui}^s}} |u\right) = \mathop \sum \limits_{j = 1}^J \mathop \sum \limits_{k = 1}^K P\left( {{w_{ui}^t}|{z_{i}^t} = j} \right)P\left( {{w_{ui}^s}|{z_i^s} = k} \right)P\left( {{z_{i}^t} = j,{z_{i}^s} = k} | u\right).
\end{equation}

Given the joint modeling of $w^t$ and $w^s$, one needs to update $\left(z^t,z^s\right)$ in a single step. The Gibbs sampling procedure can be easily derived following \citet{griffiths2004finding}:

\begin{multline}
  P\left( {z_i^t = j,z_i^s = k|w_i^t = t,w_i^s = s,{\bf{z}}_{ - i}^t,{\bf{z}}_{ - i}^s,{\bf{w}}_{ - i}^t,{\bf{w}}_{ - i}^s} \right)  \\ \propto
\frac{C_{tj}^{TJ} + \beta }{\sum\nolimits_{t' }C_{t' j}^{TJ} + T\beta }
 \times
\frac{{C_{sk}^{SK} + \gamma }}{\sum\nolimits_{s' }C_{s' k}^{SK} + S\gamma }\times\frac{C_{{u}jk}^{JK} + \alpha }{\sum\nolimits_{j'}\sum _{k'}C_{{u}j'k'}^{JK} + JK\alpha },
\end{multline}
where $C_{tj}^{TJ}$ is the number of occurrences that temporal word $t$ is assigned to topic $j$, $C_{sk}^{SK}$ is the number of occurrences that spatial word $s$ is assigned to topic $k$, and $C_{ujk}^{JK}$ is the number of words of traveler $u$ which are assigned to topic $(j,k)$. Note that the current instance $i$ is excluded in computing $C_{tj}^{TJ}$, $C_{sk}^{SK}$ and $C_{ujk}^{JK}$.

In each iteration, the Gibbs sampling will update $\left(z^t,z^s\right)$ for all LPR records sequentially. After a new of iterations, the sampling will reach stationary, and the random variables $\theta$, $\psi$ and $\varphi$ can be estimated by

\begin{equation}\label{equ:para}
\begin{split}
\theta_{ujk} &= \frac{{C_{{u}jk}^{JK} + \alpha }}{{\sum\nolimits_{j'} \sum\nolimits_{k'} {C_{{u}{j'k'}}^{JK} + {JK}\alpha } }}, \\
\psi_{tj} &= \frac{{C_{tj}^{TJ} + \beta }}{{\sum\nolimits_{t'} {C_{t'j}^{TJ} + T\beta } }}, \\
\varphi_{sk} &= \frac{{C_{sk}^{SK} + \gamma }}{{\sum\nolimits_{s'} {C_{s'k}^{SK} + S\gamma } }}.
\end{split}
\end{equation}

For an unseen traveler not in the training data (e.g., a traveler in the validation/test set), we can also apply Gibbs sampling to infer her topic composition $\hat{\theta}_u$. Given a set of training data the corresponding topic assignment of each LPR record from Gibbs sampling--$\left({{\bf{z}}_{{\rm{train}}}},{{\bf{w}}_{{\rm{train}}}}\right)$, we sample topic assignment $(\hat z_i^t,\hat z_i^s)$ for each spatiotemporal record of the unseen traveler $u$ by:

\begin{multline} \label{equ:sample_heldout}
P\left( {\hat z_i^t = j,\hat z_i^s = k|\hat w_i^t = t,\hat w_i^s = s,{\bf{\hat z}}_{ - i}^t,{\bf{\hat z}}_{ - i}^s,{\bf{\hat w}}_{ - i}^t,{\bf{\hat w}}_{ - i}^s; {{\bf{z}}_{{\rm{train}}}},{{\bf{w}}_{{\rm{train}}}}  } \right)
 \\
 \propto \frac{{C_{tj}^{TJ} + \hat C_{tj}^{TJ} + \beta }}{{\sum\nolimits_{t'} {\left[ {C_{t'j}^{TJ} + \hat C_{t'j}^{TJ}} \right]}  + T\beta }} \times \frac{{C_{sk}^{SK} + \hat C_{sk}^{SK} + \gamma }}{{\sum\nolimits_{s'} {\left[ {C_{s'k}^{SK} + \hat C_{s'k}^{SK}} \right]}  + S\gamma }} \times \frac{{\hat C_{ujk}^{JK} + \alpha }}{{\sum\nolimits_{j'} {\sum\nolimits_{k'} {\hat C_{uj'k'}^{JK}} }  + JK\alpha }},
\end{multline}
where $C_{tj}^{TJ}$ is the number of times that $t$ is assigned
to topic $j$ in the training data ($C_{sk}^{SK}$ defined in a similar way), and $\hat C_{tj}^{TJ}$ is the number of times that $t$ is assigned to topic $j$ within the test data except the current assignment ($\hat{C}_{sk}^{SK}$  defined in a similar way). Since the training results are fully utilized, the number of iterations for this inference is much smaller than that for training the original model (20 iterations are enough in our case). After performing the sampling, the topic distribution for the unseen traveler $u$ is given by:

\begin{equation}
\hat{\theta}_{ujk} = \frac{{\hat{C}_{{u}jk}^{JK} + \alpha }}{{\sum\nolimits_{j'}\sum\nolimits_{k'} {\hat{C}_{{u}j'k'}^{JK} + JK\alpha } }}.
\end{equation}

\subsection{Model Selection}

In this model, the numbers of topics for the spatial and temporal dimensions---$J$ and $K$---are hyperparameters which need to be set in advance. For model selection, we use \emph{perplexity} of a held-out validation set to evaluate models with varying $J$ and $K$. A lower perplexity over a validation set suggests better generalization performance. For this purpose, we randomly select a new set of travelers as a validation data set. Formally, perplexity of travel $u$ in the validation set is defined as the exponential of the negative normalized predictive likelihood of the validation data conditional on a trained model:

\begin{equation} \label{equ:hold_out_perplexity}
{\rm{perplexity}}\left( {{\bf{w}}_u^t,{\bf{w}}_u^s|{{\bf{w}}_{{\rm{train}}}}} \right) = \exp \left[ { - \frac{{\ln p\left( {{\bf{w}}_u^t,{\bf{w}}_u^s|{{\bf{w}}_{{\rm{train}}}}} \right)}}{{{N_u}}}} \right].
\end{equation}

Computing $p\left( {{{\bf{w}}_u^t},{{\bf{w}}_u^s}} | {\bf{w}}_{\text{train}}\right) $ is not a simple task as it involves integrals over $\theta$, $\psi$ and $\varphi$:

\begin{equation}
\label{equ:inte_perpl}
p\left({{\bf{w}}_u^t,{\bf{w}}_u^s|{{\bf{w}}_{{\rm{train}}}}} \right)
=\int {\prod\limits_{i = 1}^{{N_u}} {\left[ {\sum\limits_{j = 1}^J {\sum\limits_{k = 1}^K {\theta _{ujk}\psi _{w_{ui}^tj}\varphi _{w_{ui}^sk}} } } \right]} p\left( {\theta |{{\bf{w}}_{{\rm{train}}}}} \right)p\left( {\psi |{{\bf{w}}_{{\rm{train}}}}} \right)p\left( {\varphi |{{\bf{w}}_{{\rm{train}}}}} \right)d\theta d\psi d\varphi }.
\end{equation}

Here, we can use Monte Carlo method approximate the integral in Eq.~(\ref{equ:inte_perpl}) using $M$ point estimates from the Markov chain and then compute the average over $M$ samples:

\begin{equation} \label{equ:perp_prob}
p\left( {{\bf{w}}_u^t,{\bf{w}}_u^s|{{\bf{w}}_{{\rm{train}}}}} \right) = \frac{1}{M}\sum\limits_{m = 1}^M {\prod\limits_{i = 1}^{{N_u}} {\left[ {\sum\limits_{j = 1}^J {\sum\limits_{k = 1}^K { \theta _{ujk} \psi _{w_{ui}^tj}^m \varphi _{w_{ui}^sk}^m} } } \right]} },
\end{equation}
where $\theta _{ujk}$ are equal weights as implied by the prior distribution.

The average perplexity of a validation set to is used to score models, and a grid search can be performed to identify the best $J$ and $K$.

\subsection{Measuring Anomaly}

After model inference and selection, we obtain two sets of spatial and temporal patterns that characterizes traveler behavior of travelers in the training set. In the proposed generative model, we assume that the future travel behavior of an individual can be reconstructed by the linear combination of those basis spatial/temporal patterns. With this assumption, we consider an individual's future behavior as anomalies in the sense that future spatiotemporal records cannot be well reconstructed by the inferred latent patterns. Therefore, we consider the unpredictability of new spatiotemporal records a proxy to the degree of behavior anomaly. Here we use predictive \emph{perplexity} on an individual's future records conditional on her past behavior as a function to score behavior anomaly:

\begin{equation} \label{equ:future_perplexity}
\text{perplexity}\left( {{{\bf{\bar{w}}}_u^t},{{\bf{\bar{w}}}_u^s}} |{\bf{w}}_u^t,{\bf{w}}_u^s \right) = \exp\left[  -\frac{\ln p\left(  {{{\bf{\bar{w}}}_u^t},{{\bf{\bar{w}}}_u^s}} |{\bf{w}}_u^t,{\bf{w}}_u^s \right)}{\bar{N}_u} \right],
\end{equation}
where $\left({{{\bf{\bar{w}}}_u^t},{{\bf{\bar{w}}}_u^s}}\right)$ a set of future/new records from traveler $u$,  $\left({\bf{w}}_u^t,{\bf{w}}_u^s\right)$ represents her records in the training data, is $\bar{N}_u$ is the number of new records. The conditional probability can be approximated using $M$ Monte Carlo simulations:

\begin{equation} \label{equ:pre_perp}
p\left( {{\bf{\bar w}}_u^t,{\bf{\bar w}}_u^s|{\bf{w}}_u^t,{\bf{w}}_u^s} \right) = \frac{1}{M}\sum\limits_{m = 1}^M {\prod\limits_{i = 1}^{{{\bar N}_u}} {\left[ {\sum\limits_{j = 1}^J {\sum\limits_{k = 1}^K {\theta _{ujk}^m\psi _{\bar w_{ui}^tj}^m\varphi _{\bar w_{ui}^sk}^m} } } \right]} },
\end{equation}
where $\theta$, $\psi$ and $\varphi$ come from $M$ stationary Gibbs chains during model inference (see Eq.~(\ref{equ:para})).

Note that Eq.~(\ref{equ:hold_out_perplexity}) computes the perplexity of a held-out traveler, while on the other hand Eq.~(\ref{equ:future_perplexity}) measures the model's performance in predicting future records of the same traveler $u$ given her past records. Thus, this value gives insights about whether a person is predictive or not. A high perplexity indicates that it is difficult to use historical patterns to predict the travel behavior on the following days. A low perplexity indicates the learned patterns can be perfectly used to model future behavior.

\section{Numerical Experiment}  \label{sec:experiments}

In this section, we present a case study using a four-week LPR data set in Guangzhou, China (see Section~\ref{sec:data}). As previously mentioned, the sample data set we use consists of LPR transactions from 2200 unique license plate numbers. Here we consider each unique number a unique user/traveler. The average number of records per traveler is 1331. We divide these 2200 travelers into two groups: 2000 as a training set (for pattern discovery and anomaly detection), and the rest 200 as a validation set (for model selection with regard to hyperparameters $J$ and $K$).

In the training data (2000 travelers), records from the first three weeks are used for pattern discovery and records from the last week are used to quantify the predictive perplexity (degree of anomalies). For hyperparameters in the LDA model, we set $\alpha=0.01$ and $\beta=0.01$. We set $M=10$ for the Monte Carlo integration. We perform model inference on a grid with $J=8, 10, 12, 14$ and $K=15, 20, 25, 30$, and compute the average perplexity on the 500 travelers in the validation set. The best model is obtained when $J=10$ and $K=25$, with an average validation perplexity of $2.495\times 10^3$.

\subsection{Interpreting Latent Factors}

\subsubsection{Temporal factors}

\begin{figure}[!ht]
  \centering
  \includegraphics[scale=0.5]{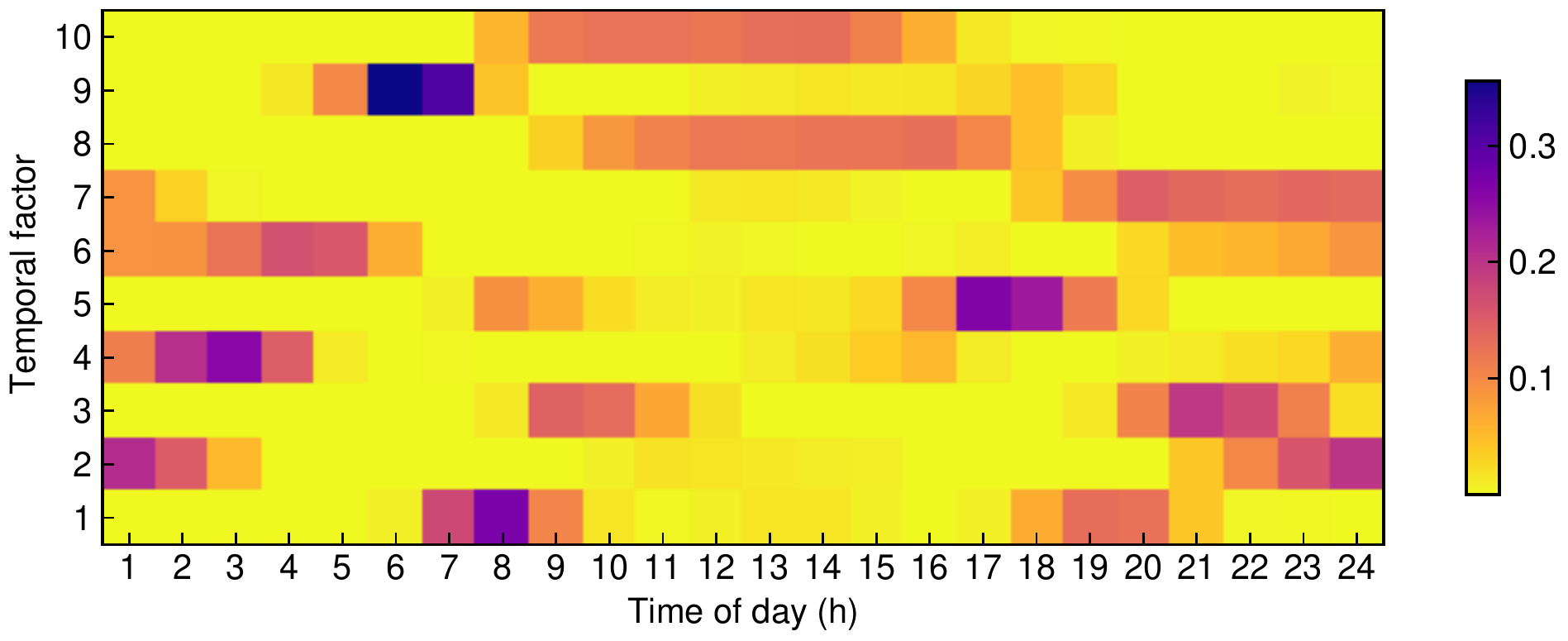}
   \caption{The distribution of $\psi_{j}$ for the $J=10$ temporal factors.}\label{fig:temporal_patterns}
\end{figure}

Figure~\ref{fig:temporal_patterns} shows the distribution of the 10 temporal factors $\psi_{j}$. As can be seen, each $\psi_{j}$ characterizes very unique patterns from each other. For example, $\psi_{1}$ mainly concentrates on 7-9 am and 6-9 pm, which typically capture work-related activities. $\psi_{2}$ mainly covers late-night patterns from 9 pm to 3 am. $\psi_{8}$ and $\psi_{10}$ shows two similar temporal patterns for non-peak hours from 10 am to 4 pm.

\subsubsection{Spatial factors}

\begin{figure}[!ht]
  \centering
  \includegraphics[width=\linewidth]{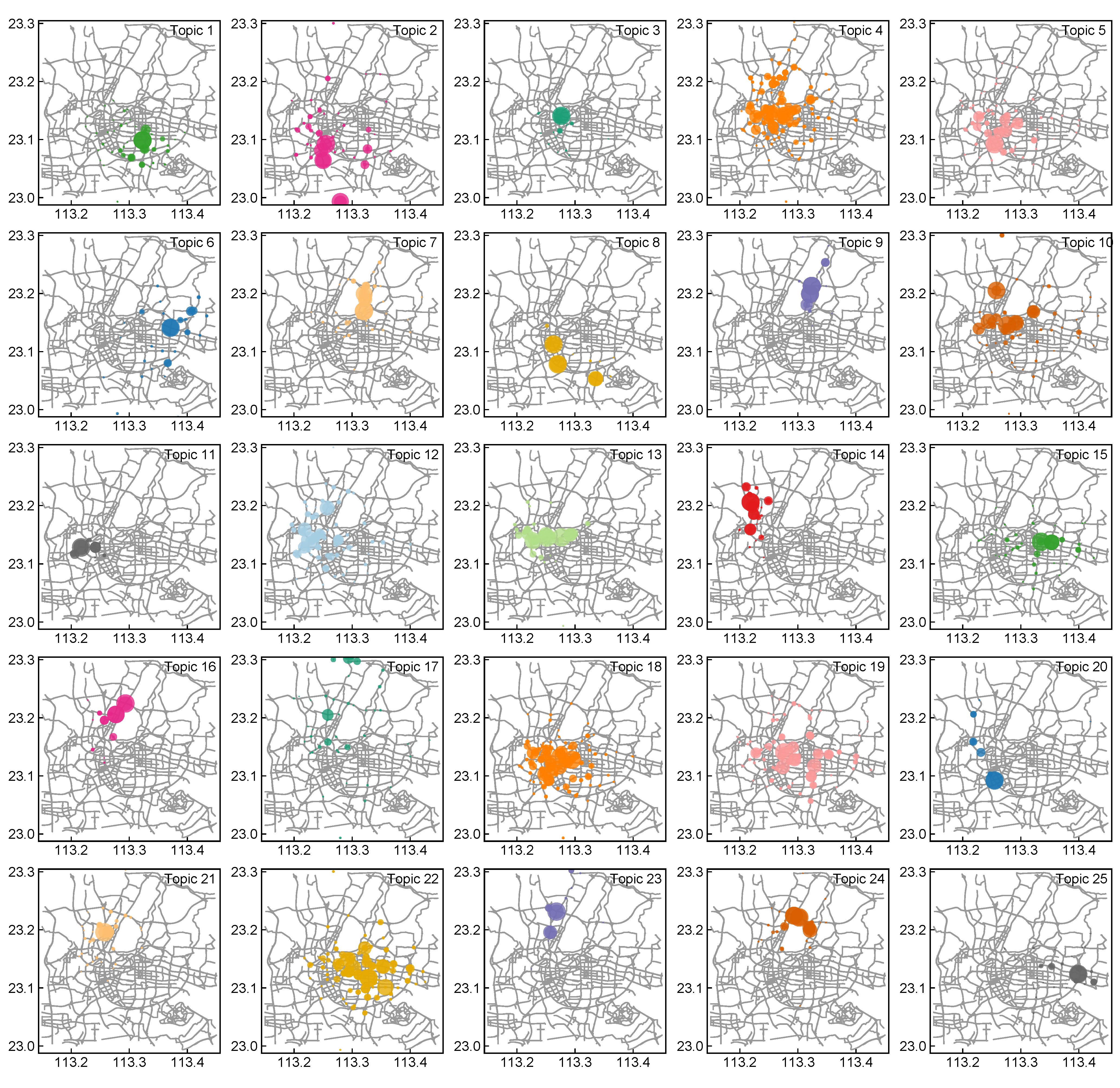}
   \caption{The distribution of $\phi_{k}$ for the $K=25$ spatial factors. Size of node/detector corresponds to the probability $\phi_{sk}$.}\label{fig:spatial_patterns}
\end{figure}

On the spatial dimension, we also uncover meaningful patterns with clear clustering patterns and spatial consistencies, with each topic $\phi_{k}$ concentrating on certain area of the city (see Figure~\ref{fig:spatial_patterns}). Note that modeling training only uses the labels of detectors, instead of their exact locations. Thus, the spatial consistency shown in Figure~\ref{fig:spatial_patterns} further verifies the effectiveness of our model. It should be noted that topics that look similar in Figure~\ref{fig:spatial_patterns} (e.g.,  $\phi_{4}$ and  $\phi_{18}$) may represent very different distributions, due to the directional properties of the transportation networks. Some topics---such as $\phi_{3}$, $\phi_{8}$, and $\phi_{14}$---show very unique distributions with little overlap with other topics.

\subsection{Similarity in Travel Behavior}

\begin{figure}[!ht]
  \centering
  \includegraphics[scale=0.5]{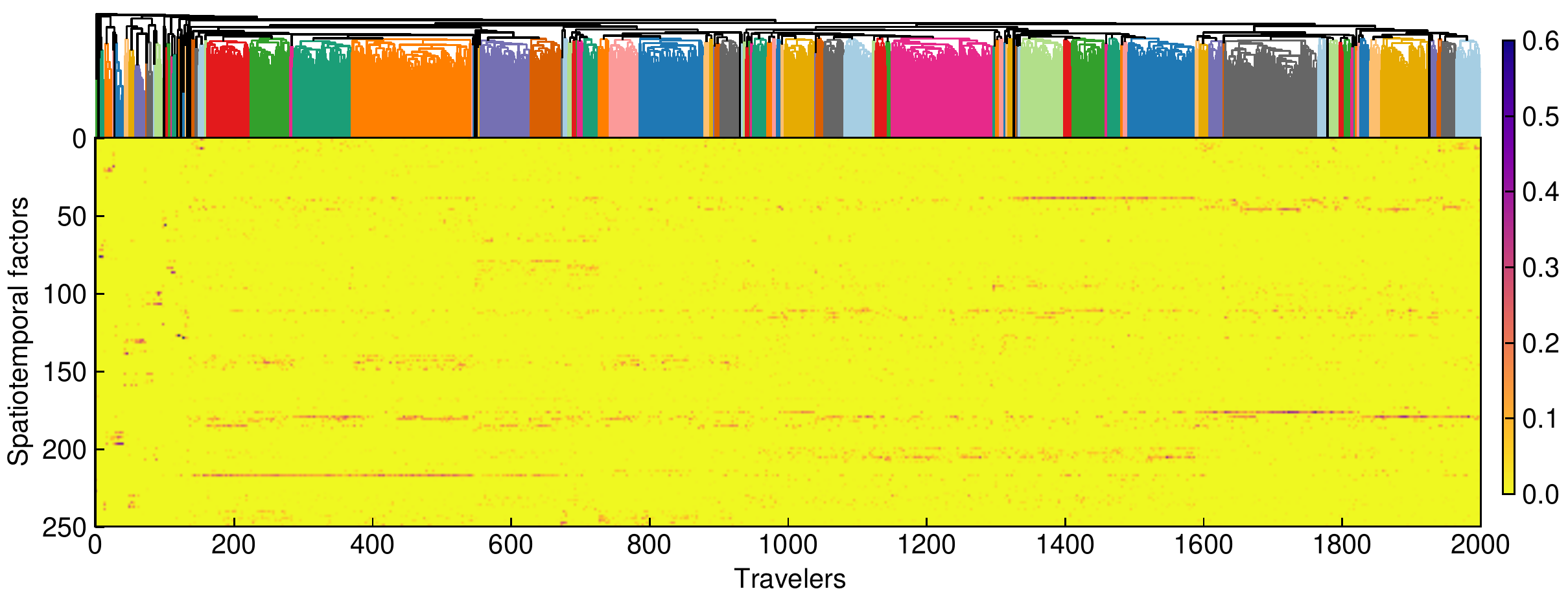}
  \caption{Hierarchical clustering of 2000 travelers based on the Jensen-Shannon distance (square root of the Jensen-Shannon divergence $\sqrt{JSD\left(u,v\right)}$) and the average linkage function.}\label{fig:clustering}
\end{figure}

As the model factorizes the complex data into a weighted combination of a few latent patterns, it gives us a compact representation of an individual's travel behavior as the topic distribution matrix $\theta$. In other words, we summarize the complex behavior of a traveler into a low-dimensional latent representation. Based on this representation, we can measure the similarity between two travelers $u$ and $v$ using distance measures. Since $\theta_u$ and $\theta_v$ are probability distributions, we quantify their difference using the Jensen-Shannon divergence:
\begin{equation}
JSD\left(u,v\right) = \frac{1}{2}\sum\limits_{j = 1}^J {\sum\limits_{k = 1}^K {\left[ {{\theta _{ujk}}\log \frac{{{\theta _{ujk}}}}{{{{\bar \theta }_{jk}}}} + {\theta _{vjk}}\log \frac{{{\theta _{vjk}}}}{{{{\bar \theta }_{jk}}}}} \right]} },
\end{equation}
where ${{\bar \theta }_{jk}}=\frac{1}{2}\left(\theta_{ujk}+\theta_{vjk}\right)$.

\begin{figure}[!ht]
  \centering
  \includegraphics[scale=0.5]{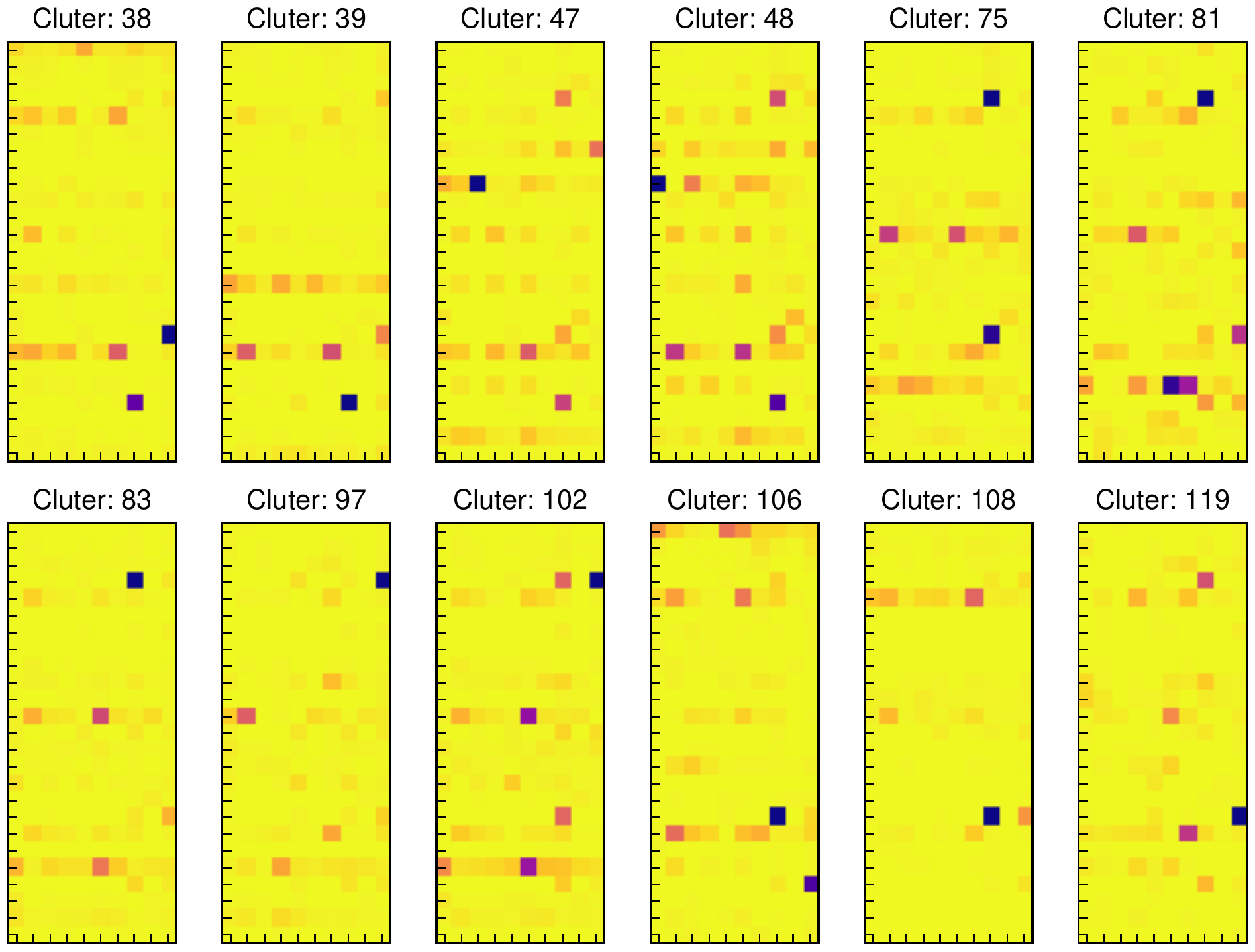}
   \caption{Average topic distribution $\theta$ of travelers in 12 randomly selected clusters.}\label{fig:example_clusters}
\end{figure}

By working with latent representation $\theta$ instead of comparing the raw spatiotemporal records, we significantly reduce the complexity in quantifying the travelers' similarity. As an illustration, we select 500 travelers randomly from the training set and compute pairwise Jensen-Shannon divergence based on their topic distribution $\theta$.  We run hierarchical clustering by using the average linkage function and $\sqrt{JSD}$ as a distance metric. The dendrogram in Figure~\ref{fig:clustering} shows the result of hierarchical clustering and the matrix represents the topic strength $\theta$. The result further demonstrates the effectiveness of our model in identifying travelers with similar behavior patterns efficiently.

Different from applying k-mean clustering based on specific pre-defined variables as in \citet{chen2017clustering}, the individual topic distribution $\theta$ provides a natural way to summarize complex travel history data and measure traveler similarity using probability measures. Travelers in the same cluster have small divergence, and thus they share similar daily routine patterns. Such information could be very useful in designing shared transport services (e.g., carpooling, transit network planning, and customized bus services) and identifying group travel patterns. Figure~\ref{fig:example_clusters} shows some the averaged topic distribution ($\theta$) of 12 randomly selected clusters obtained by the above hierarchical clustering. As we can see, most clusters shows a parsimonious and sparse pattern, with most of the weight concentrating a only a few $(j,k)$ combinations. This indicates that travelers in the same cluster do exhibit very similar travel patterns in terms of both time and location. This further suggests the strong regularity in collective travel behavior patterns.

\subsection{Anomaly Detection for Travelers}

\begin{figure}[!ht]
\centering
\includegraphics[scale=0.5]{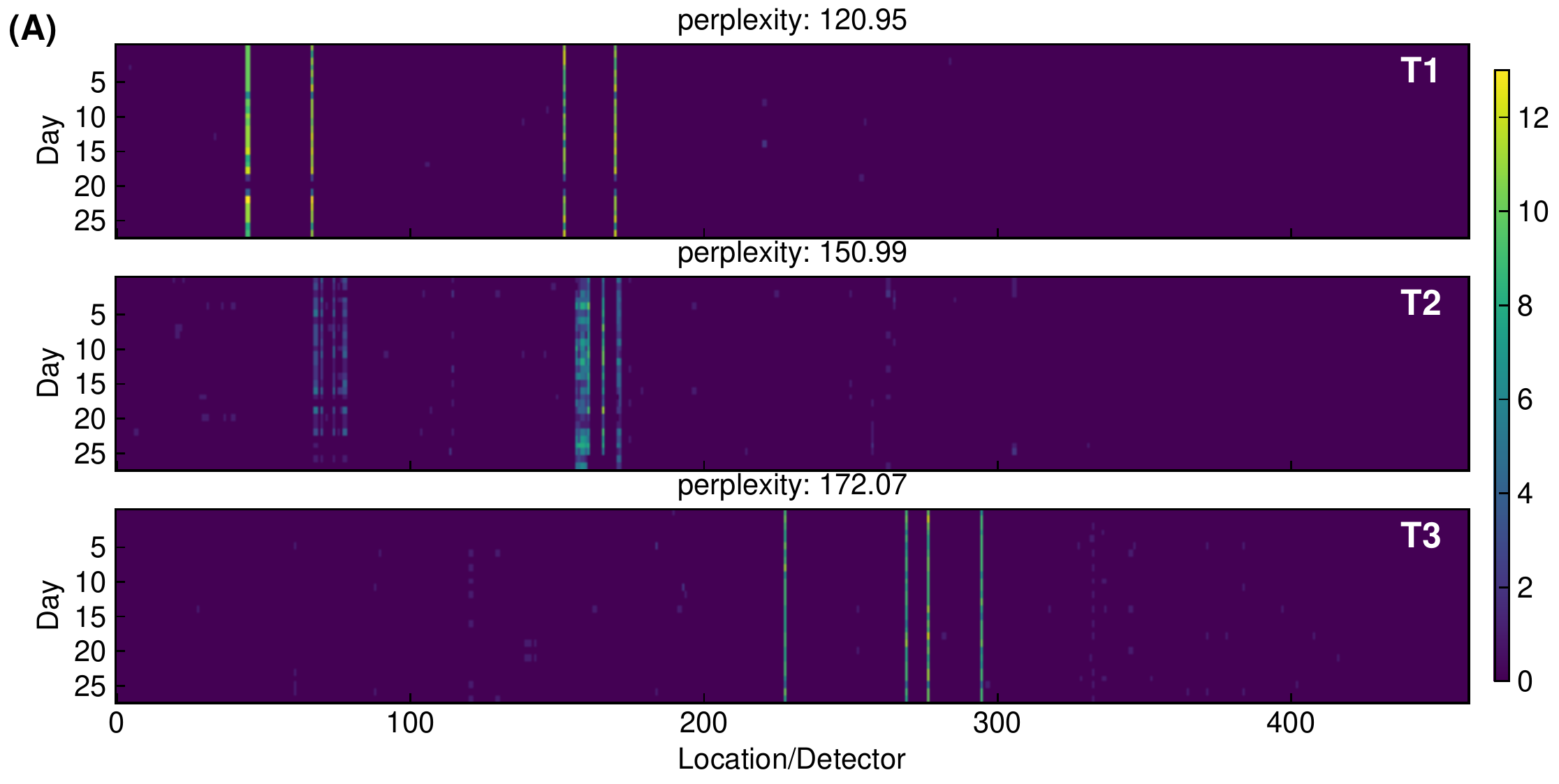}
\includegraphics[scale=0.5]{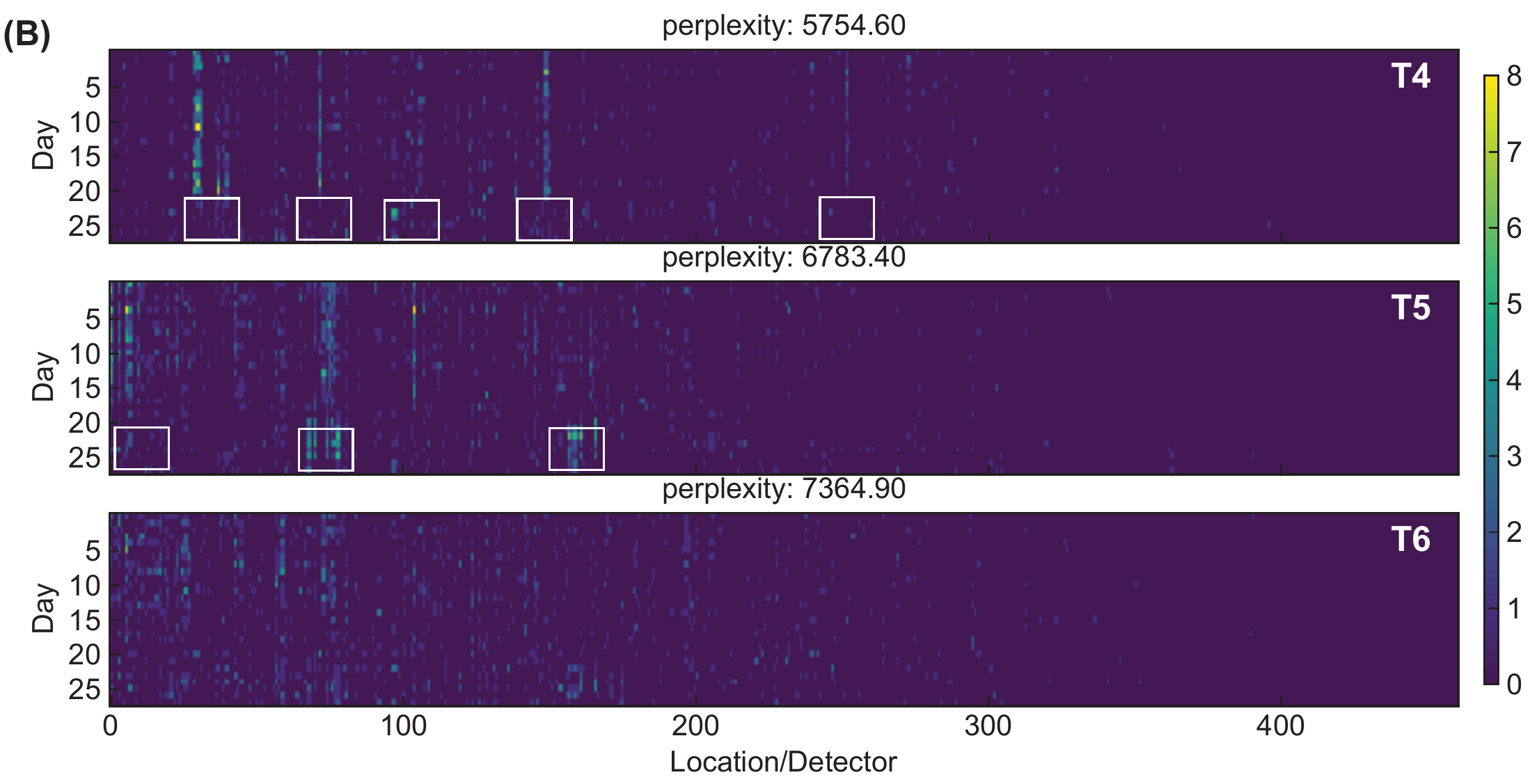}
\caption{The detector visiting frequencies across the four weeks for selected travelers. (A) Three travelers (T1, T2, T3) with lowest perplexity. (b) Three travelers (T4, T5, T6) with highest perplexity.}
\label{fig:six_travelers}
\end{figure}

After training the model using the first three week's data from the 2000 travelers, we apply the inferred model on their records in the last week to compute predictive perplexity using Eq.~(\ref{equ:future_perplexity}). As mentioned, the predictive perplexity is a proxy to the degree of behavior anomaly, and it allows us to rank travelers according to their anomaly. In computing the predictive perplexity, we use samples from $M=10$ Markov chains for Monte Carlo integration. The average predictive perplexity of these 2000 travelers is $3.503\times 10^3$ (min $120.95$ and max $9.564\times 10^3$). The value allows us to distinguish routine travelers (with low perplexity) from random explorers (with high perplexity). To better understand how perplexity interacts with behavior anomaly, we select three travelers with the lowest perplexity (T1, T2, T3) and three travelers with the highest perplexity (T4, T5, T6) as examples. Figure~\ref{fig:six_travelers} shows the location/detector visiting frequency for the six travelers in each day. Panel (A) shows three travelers with the lowest perplexity. These travelers exhibit stable/regular travel patterns across all the four weeks. And their behavior in the fourth week can be well reproduced by their past travel records. In other words, their behavior is of high regularity, with routine trips visiting the same location and the same time from day to day.

Panel (B) shows three travelers among those with the highest perplexity values. This panel shows that essentially there exist many factors leading to a high predictive perplexity. On the one hand, the traveler could be a random explorer (e.g., a taxi driver), whose travel behavior does not show a repetitive pattern in the LPR data clearly (e.g., T6). For these travelers, their past behavior does not provide much information about the modeling of their future behavior, and thus it is difficult to perform prediction. On the other hand, a high perplexity might result from the case where one's travel behavior suddenly changes as described in  \citet{zhao2018detecting} (e.g., T4 and T5). These changes might be attributed to factors such as change of home/work location, seasonal effect (e.g., from semester to vacation for students), and one's car being borrowed by others. In these cases, it also becomes difficult to predict one's behavior in the last week based on travel patterns learned from the first three weeks, even though she has routine travel behavior in the first three weeks.

\section{Conclusion and Discussion} \label{sec:conclusion}

The paper investigates the problem of pattern discovery and anomaly detection on individual travel behavior data. The overall objective is to handle large amounts of complex spatiotemporal data efficiently and effectively. We present a probabilistic framework to model spatiotemporal travel behavior data. In detail, we use a two-dimensional LDA model to reproduce complex travel behavior from a linear combination of meaningful patterns on spatial and temporal dimensions. An essential feature of this model is to allow spatial and temporal patterns to interact with each other by using a two-dimensional topic distribution simplex; thus, it can capture complex spatiotemporal data by using only a few patterns. We implement an efficient collapsed Gibbs sampling algorithm for model inference. Based on the assumption that the model characterizes the ``normal'' data generation process, it can be further used to detect anomalies in individual travel behavior. We define the degree of behavior anomaly for an individual as the ``unpredictability'' of her future activity under the trained model. Specifically, we use predictive perplexity to measure an individual's travel behavior anomalousness quantitatively. As a numerical experiment, we apply this model to four week's LPR data and show the performance of this framework with a traveler clustering example and an anomaly ranking example. This type of anomaly detection applications will provide insights for traffic monitoring, law enforcement, individual behavior profiling, and even insurance pricing since the model clearly distinguishes routine travelers from random explorers.

There are many directions for future research: (1) Individual behavior patterns (topic distribution) may change gradually over time. This type of natural change is not considered in the current model. For this, one may include another ``day of week'' dimension to take the behavior shifts over time into account; (2) Another research direction is to extend this model for travel behavior prediction with accurate timestamps. The current model is built on a ``bag-of-word'' assumption, and all records are considered exchangeable; however, time order is crucial for synthetic activity chain generation. To create a realistic order of spatiotemporal points, we may integrate a hidden Markov model to characterize the sequential property of records, such as in \citet{fan2016collaborative} and \citet{yin2018generative}; (3) The current model performs pattern discovery and anomaly detection in an unsupervised way. In fact, we can show the results to field expert and gather feedback from them, and this will allow us to further employ supervised methods to improve the quantity of the model; (4) Additional information such as personal attributes if available (e.g., type of vehicle, age/gender of drivers) of the traveler can be also integrated into this model as labels to support supervised training; (5) This paper currently model travel behavior on a daily basis with ``time of day'', while real-world behavior also exhibits strong ``day of week'' patterns. Our model can be further developed by introducing additional dimensions such as ``day of week''.

\section*{Acknowledgement}
This research is supported by the Natural Sciences and Engineering Research Council (NSERC) of Canada, Mitacs Canada, Canada Foundation for Innovation, and Fundway Technology Inc. The MATLAB code for this project is available at \url{https://github.com/lijunsun/travel_behavior_anomaly}.

\end{document}